\documentclass[english,useAMS,usenatbib]{mn2e} 
\usepackage{lmodern}

\usepackage[T1]{fontenc}
\usepackage[utf8]{inputenc}
\usepackage{color}
\usepackage[usenames,dvipsnames]{xcolor}
\usepackage{amsmath}
\usepackage{amssymb}
\usepackage{graphicx}
\usepackage{esint}
\usepackage{placeins}
\usepackage[authoryear]{natbib}

\usepackage{hyperref}
\definecolor{darkgreen}{rgb}{0.1,0.6,0.7}
\hypersetup{colorlinks = true,urlcolor=blue,citecolor=Blue}

\makeatletter
\usepackage{lmodern}\usepackage{epstopdf}

\let\jnl@style=\rm
\def\ref@jnl#1{{\jnl@style#1}}

\def\aj{\ref@jnl{AJ}}                   
\def\actaa{\ref@jnl{Acta Astron.}}      
\def\araa{\ref@jnl{ARA\&A}}             
\def\apj{\ref@jnl{ApJ}}                 
\def\apjl{\ref@jnl{ApJ}}                
\def\apjs{\ref@jnl{ApJS}}               
\def\ao{\ref@jnl{Appl.~Opt.}}           
\def\apss{\ref@jnl{Ap\&SS}}             
\def\aap{\ref@jnl{A\&A}}                
\def\aapr{\ref@jnl{A\&A~Rev.}}          
\def\aaps{\ref@jnl{A\&AS}}              
\def\azh{\ref@jnl{AZh}}                 
\def\baas{\ref@jnl{BAAS}}               
\def\bac{\ref@jnl{Bull. astr. Inst. Czechosl.}}
\def\caa{\ref@jnl{Chinese Astron. Astrophys.}}
\def\cjaa{\ref@jnl{Chinese J. Astron. Astrophys.}}
\def\icarus{\ref@jnl{Icarus}}           
\def\jcap{\ref@jnl{J. Cosmology Astropart. Phys.}}
\def\jrasc{\ref@jnl{JRASC}}             
\def\memras{\ref@jnl{MmRAS}}            
\def\mnras{\ref@jnl{MNRAS}}             
\def\na{\ref@jnl{New A}}                
\def\nar{\ref@jnl{New A Rev.}}          
\def\pra{\ref@jnl{Phys.~Rev.~A}}        
\def\prb{\ref@jnl{Phys.~Rev.~B}}        
\def\prc{\ref@jnl{Phys.~Rev.~C}}        
\def\prd{\ref@jnl{Phys.~Rev.~D}}        
\def\pre{\ref@jnl{Phys.~Rev.~E}}        
\def\prl{\ref@jnl{Phys.~Rev.~Lett.}}    
\def\pasa{\ref@jnl{PASA}}               
\def\pasp{\ref@jnl{PASP}}               
\def\pasj{\ref@jnl{PASJ}}               
\def\rmxaa{\ref@jnl{Rev. Mexicana Astron. Astrofis.}}%
\def\qjras{\ref@jnl{QJRAS}}             
\def\skytel{\ref@jnl{S\&T}}             
\def\solphys{\ref@jnl{Sol.~Phys.}}      
\def\sovast{\ref@jnl{Soviet~Ast.}}      
\def\ssr{\ref@jnl{Space~Sci.~Rev.}}     
\def\zap{\ref@jnl{ZAp}}                 
\def\nat{\ref@jnl{Nature}}              
\def\iaucirc{\ref@jnl{IAU~Circ.}}       
\def\aplett{\ref@jnl{Astrophys.~Lett.}} 
\def\apspr{\ref@jnl{Astrophys.~Space~Phys.~Res.}}
\def\bain{\ref@jnl{Bull.~Astron.~Inst.~Netherlands}} 
\def\fcp{\ref@jnl{Fund.~Cosmic~Phys.}}  
\def\gca{\ref@jnl{Geochim.~Cosmochim.~Acta}}   
\def\grl{\ref@jnl{Geophys.~Res.~Lett.}} 
\def\jcp{\ref@jnl{J.~Chem.~Phys.}}      
\def\jgr{\ref@jnl{J.~Geophys.~Res.}}    
\def\jqsrt{\ref@jnl{J.~Quant.~Spec.~Radiat.~Transf.}}
\def\memsai{\ref@jnl{Mem.~Soc.~Astron.~Italiana}}
\def\nphysa{\ref@jnl{Nucl.~Phys.~A}}   
\def\physrep{\ref@jnl{Phys.~Rep.}}   
\def\physscr{\ref@jnl{Phys.~Scr}}   
\def\planss{\ref@jnl{Planet.~Space~Sci.}}   
\def\procspie{\ref@jnl{Proc.~SPIE}}   

\newcommand{\boS}{{\ensuremath{\sf{S}}}}
\newcommand{\boC}{{\ensuremath{\sf{C}}}}

\newcommand{\sfS}{\ensuremath{{\sf{S}}}}

\newcommand{\mathd}{\ensuremath{\mathrm{d}}}

\newcommand{\calP}{\ensuremath{\mathcal{P}}}

\newcommand{\calH}{\ensuremath{\mathcal{H}}}

\newcommand{\calG}{\ensuremath{\mathcal{G}}}
\newcommand{\calC}{\ensuremath{\mathcal{C}}}
\newcommand{\calD}{\ensuremath{\mathcal{D}}}

\newcommand{\fatx}{\ensuremath{\boldsymbol{x}}}
\newcommand{\fatk}{\ensuremath{\boldsymbol{k}}}

\title[Non-Gaussian likelihoods]{On the insufficiency of arbitrarily precise covariance matrices: non-Gaussian weak lensing likelihoods}

\author[Sellentin \& Heavens]{Elena Sellentin$^{1,2}$, Alan F. Heavens$^{2}$\\
$^{1}$Département de Physique Théorique, Université de Genève, 24 Quai Ernest-Ansermet, CH-1211 Genève, Switzerland\\
$^{2}$Imperial Centre for Inference and Cosmology (ICIC), Imperial College, Blackett Laboratory, Prince Consort Road, London SW7 2AZ, U.K.
}

\makeatother

\usepackage{babel}
\begin{document}

\date{Accepted 30 BC. Received 800 AD; in original form 10,000 BC}

\maketitle
\pagerange{\pageref{firstpage}--\pageref{lastpage}} \pubyear{2016}

\label{firstpage} 
\begin{abstract}
We investigate whether a Gaussian likelihood, as routinely assumed in the analysis of cosmological data, is supported by simulated survey data. We define test statistics, based on a novel method that first destroys Gaussian correlations in a dataset, and then measures the non-Gaussian correlations that remain. This procedure flags pairs of datapoints which depend on each other in a non-Gaussian fashion, and thereby identifies where the assumption of a Gaussian likelihood breaks down. Using this diagnostic, we find that non-Gaussian correlations in the CFHTLenS cosmic shear correlation functions are significant. 
With a simple exclusion of the most contaminated datapoints, the posterior for $s_8$ is shifted without broadening, but we find no significant reduction in the tension with $s_8$ derived from Planck Cosmic Microwave Background data. However, we also show that the one-point distributions of the correlation statistics are noticeably skewed, such that sound weak lensing data sets are intrinsically likely to lead to a systematically low lensing amplitude being inferred. The detected 
non-Gaussianities get larger with increasing angular scale such that for future wide-angle surveys such as Euclid or LSST, with their very small statistical errors, the large-scale modes are expected to be increasingly affected. The shifts in posteriors may then not be negligible and we recommend that these diagnostic tests be run as part of future analyses.
\end{abstract}


\section{Introduction}
In the past four years, major scientific effort has gone into the estimation of covariance matrices for the large-scale structure observations, see e.g.\ \citet{Hartlap,TJ,TJK,Dodelson,SH15,SH17} and references therein. A remaining question in this context is however whether obtaining an arbitrarily precise covariance matrix is sufficient, or whether non-Gaussian correlations between the data points exist, and should be accounted for.

This question is especially of importance for cosmological observables which are derived from the cosmic large-scale structure (LSS). For upcoming experiments like Euclid \citep{Euclid}, LSST \citep{Synergies}, but also for current experiments like CFHTLenS \citep{Joudaki16, Heymans}, KiDS \citep{KiDS}, DES \citep{DES}, and eBOSS \citep{eBOSS}, the observables are galaxies who trace the underlying dark-matter fields in a biased way. Associated observables can either be weak lensing or galaxy clustering counts or other observables like the abundance of extremely massive galaxy clusters, or peak counts in weak-lensing maps.

Common to all of these observables is that they arise from an underlying matter field, which has undergone gravitational evolution for the past 13 billion years, and thereby built up structures whose statistics deviate decisively from Gaussianity. It is therefore likely that estimators derived from such non-Gaussian fields follow non-Gaussian distributions themselves. This would imply that datasets from the low-redshift Universe are non-Gaussian multivariate random variables, and if we nonetheless use a Gaussian likelihood when measuring cosmological parameters, then the inappropriate shape of the likelihood will be a source of systematics. As the likelihood weights the different datapoints and determines their correlations, it is essentially impossible to predict how choosing an inadequate likelihood affects the final parameter constraints. It is however inevitable that the final posterior of parameters will be biased if the wrong likelihood for the data is supposed.

The usual counter-argument envoked to appease worries about non-Gaussian likelihoods is the central limit theorem (CLT). 
It ensures that if enough random variables $x_i$ are drawn from an unspecified distribution $\calD(x)$ of finite variance, then the distribution of their mean $\bar{x} = 1/N \sum_{i=1}^N x_i$ tends towards a Gaussian distribution, for $N \rightarrow \infty$. This is because the magnitude of higher-order cumulants is reduced in the averaging process.

On the other hand, it is however also true, that any non-linear function of a Gaussian random variable will automatically  be non-Gaussianly distributed. So it is a priori not clear which of these two effects dominates: the Gaussianization due to the CLT, or the de-Gaussianization due to non-linear functions, of which non-linear structure growth is only one example. Another possibility is the presence of systematic effects whose distributions may well not be Gaussian.

The construction of a multi-dimensional non-Gaussian likelihood function in general would be a difficult, if not impossible task.  In this paper, we therefore simply measure whether a selected variety of cosmological estimators follows a Gaussian distribution or not. One way to test for non-Gaussianity is to calculate higher order cumulants, such as the bi- or trispectrum to a powerspectrum. Such calculations probe however only the first few orders beyond the Gaussian approximation, although any non-Gaussian distribution automatically has infinitely many non-zero higher-order cumulants. Here we will therefore provide a suite of non-Gaussian tests that are sensitive with respect to all higher orders at once, because it uses properties of the entire \emph{likelihood}-shape, rather than just properties of its moments and cumulants. This is akin to our previously presented non-Gaussian likelihood expansion \citep{DALI, DALII}, where strong non-Gaussianities are also captured by a deformation of the likelihood, rather than by the inclusion of higher-order cumulants. 

Here, we compute three matrices $\boS^+, \boS^*, \boS^\div$ which have the same structure as a covariance matrix, i.e. the $(i,j)$-element of any matrix $\boS^{\{+,*,\div\}}$ measures the coupling between the $i$th and $j$th element of a $d$-dimensional datavector $\fatx$. However, in contrast to the covariance matrix which measures the Gaussian correlation between the two data points, these matrices will measure the non-Gaussian correlations and we hence refer to them as \emph{trans}-covariance matrices, where we borrow the Latin meaning of the prefix `trans' to indicate everything that goes beyond the Gaussian level. For a truly Gaussian dataset, only covariances exist and all trans-covariances will then be statistically compatible with zero. We will however show that this is not the case for available cosmological datasets, and the non-zero elements of the trans-covariance matrices succeed in flagging data points with non-Gaussian statistics in cosmological datasets. For the future generation of large-scale structure observations, we recommend that this suite is run on the simulations from which the covariance is usually measured, as it provides valuable insights into where a Gaussian likelihood has to be distrusted.

\begin{figure*}
\includegraphics[width=\textwidth]{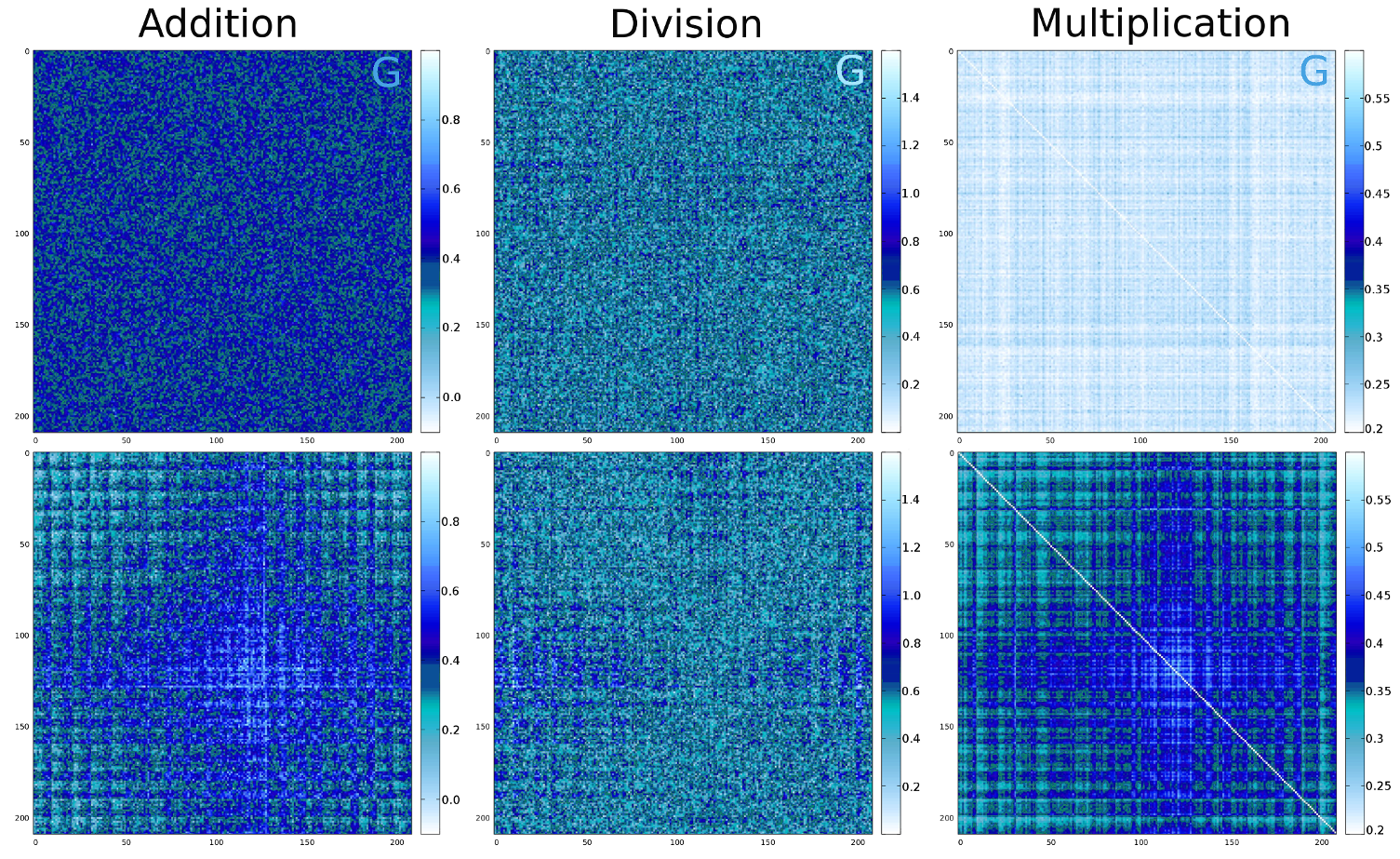} 
\caption{ $\boS^+,\boS^\div$ and $\boS^*$ for the Gaussian calibration runs (top row) and the local model of non-Gaussianity with $f_{\rm nl = 0.5}$ (bottom row). Each of the depicted matrices is structured like a covariance matrix, meaning the data vector runs along the two axes of the plot. The displayed matrices depict however non-Gaussian correlations between two elements of the data vector, whereas a plot of the covariance matrix would depict the Gaussian covariance between the two elements. The top row depicts the effect of shotnoise when computing the matrices for Gaussian calibration data. The bottom row depicts the trans-covariance matrices for the $f_{\rm nl}$ model, in the same colour scheme as their respective calibration runs above. The relative difference between calibration runs and test runs for the $f_{\rm nl}$ model clearly indicates the presence of non-Gaussian correlations between the datapoints. Essentially the entire $f_{\rm nl}$ dataset is contaminated with non-Gaussian correlations, as is to be expected from the model. Smaller values of $f_{\rm nl}$ can be detected by increasing the number of simulations.}
\label{fnlplot}
\end{figure*} 
\section{Consequences of Gaussianity}
\label{Sec:Sec1}
Let us assume we have an estimator of a cosmological observable, which could for example be the power spectrum $P(k)$, a correlation function $\xi(\theta)$ or a spherical harmonic $C_{\ell}$ of some field. Statistically speaking, these are data, and hence multivariate random variables, which we will denote as $\fatx$. In virtually all standard cosmological analyses it is assumed that $\fatx$ follows a Gaussian likelihood when explaining the datasets with a parametric model for structure-growth. We now wish to test whether the Gaussian assumption is justified. In order to do so, we simply work out consequences that must be true if the Gaussian assumption is valid. 

The most agnostic tests for two supposedly Gaussian random variables is to check their behaviour under the three arithmetic operations of addition, multiplication and division. For Gaussian random variables, the following testable statements hold.

If $x_i$ and $x_j$ are independently drawn from
\begin{equation*}
 x_i \sim \calG(0,1), \ \ \ \ \ x_j \sim \calG(0,1),
\end{equation*}
then their sum follows a Gaussian distribution of variance 2,
\begin{equation}
 x_i + x_j \sim \calG(0,2).
 \label{add}
\end{equation}
Their ratio follows the standard Cauchy distribution
\begin{equation}
 y = \frac{x_i}{x_j} \sim \calC(y) = \frac{1}{\pi (1+y^2)},
 \label{ratio}
\end{equation}
and their product follows the distribution of two linearly superposed $\chi^2$-random variables with one degree of freedom. The latter arises because the product $x_ix_j$ can be rewritten as a sum of squares
\begin{equation}
 x_ix_j = \frac{1}{4} \left[  (x_i +x_j)^2 - (x_i -x_j)^2 \right],
 \label{prod}
\end{equation}
where each of the summands is a squared Gaussian random variable, and therefore a $\chi^2$-variable with one degree of freedom. For ease of notation we shall denote this distribution as $\calP$ in the following. It can easily be sampled, but does not seem to have a name.

The above three statements Eq.~(\ref{add}), Eq.~(\ref{ratio}) and Eq.~(\ref{prod}) can be tested if many statistically iid realizations of a random variable are available. As we currently estimate covariance matrices for cosmological observables from simulations, the needed random samples are available, and we will conduct these test in the following for the simulations which we had access to. Note, that since each of these tests utilizes the full distribution, these tests react to all potentially present higher-order cumulants, instead of just the lowest order ones.

\subsection{Testing procedure}
In order to test whether a given cosmological dataset is drawn from a multivariate Gaussian distribution, we proceed as follows. Let $\fatx$ be a $d$-dimensional cosmological dataset. We use the $N_s$ statistically independent realizations $\fatx_i$ from simulations which were originally run to compute the covariance matrix
\begin{equation}
 \boC = \frac{1}{N_s -1} \sum_{i = 1}^{N_s} (\fatx_i - \bar{\fatx})(\fatx_i - \bar{\fatx})^T.
\end{equation}
Let us denote the $d$ elements of the $i$th datavector as $x_i^e$, where $i \in [1,N_s]$ and $e \in [1,d]$. 
In order to test whether any two elements $x_i^e$ and $x_i^f$ of the datavector possess a non-Gaussian correlation, we first of all destroy their Gaussian correlation (their covariance) by a mean-subtraction and a whitening step which involves either diagonalizing the full covariance matrix $\boC$, or diagonalizing the two-dimensional submatrix 
\begin{equation}
 \boC_{2 \times 2} = 
 \begin{pmatrix}
  {\rm Cov(e,e) } & {\rm Cov(e,f) }\\
  {\rm Cov(f,e) } & {\rm Cov(f,f) } \\
 \end{pmatrix}.
 \label{miniwhite}
\end{equation}
If the original dataset was indeed Gaussian distributed, then both whitening procedures now must have destroyed all correlations between the data points, such that the prerequisites for the tests Eq.~(\ref{add}), Eq.~(\ref{ratio}), Eq.~(\ref{prod}) should be fulfilled. However, if non-Gaussian correlations remain, then the prerequisites are not fulfilled and the whitened datapoints will fail the following tests. To be precise, we will here always use the whitening procedure of Eq.~(\ref{miniwhite}), i.e. use the $2\times2$ covariance matrix of two data points for Cholesky whitening. Singling out the $2\times2$ submatrix has the advantage that we know which data points any eventually detected non-Gaussian correlation refers to. If we would always whiten the entire data set at once, the non-Gaussian correlations would be detected in the principal components, which are superpositions of the individual data points. This may be of interest as well, but we here prefer to work on the basis of individual data points.

In order to test whether Eq.~(\ref{add}) holds, we compute for each pair $x^e_i,x_i^f$ with $e \neq f$ the sum
\begin{equation}
 s_i^{e,f} = x_i^e + x_i^f.
\end{equation}
For each pair $(e,f)$ this will produce $N_s$ samples of their sum $s_i$. These samples are then distributed onto the $B$ bins $\calH_b$ of a histogram. If the tested dataset was indeed Gaussian, then for $N_s \rightarrow \infty$ and $B \rightarrow \infty$ this histogram will tend to the Gaussian distribution $\calG(0,2)$. We have considered a number of measures of deviation from the expected distributions for Gaussian data, including the Kullback-Leibler divergence (which we find to be overly sensitive to the tails of the distribution for our purposes), and the $L_1$ norm, which performs similarly to the measure we chose, which is the total quadratic distance of the histogram bins to the Gaussian expectation. This is the so-called MISE error, as familiar from density estimation techniques (Mean Integrated Squared Error)
\begin{equation}
\frac{1}{B} \sum_{b = 1}^B [ \calH_b(s_i^{e,f}) - \calG(0,2)  ]^2 =: \boS^+_{e,f}.
 \label{Splus}
\end{equation}
The above MISE error is a measure for the deviation between the expected distribution of the sum for Gaussian random samples, and the emerging distribution of the sum for the tested samples. We compute $S^+_{e,f}$ for any pair of data points. This builds up the matrix $\boS^+$ which will then be a summary of the non-Gaussian couplings between the data points. The total non-Gaussian contamination of the $e$th datapoint is then the sum over a column in the trans-covariance matrices
\begin{equation}
 \epsilon^{\rm tot,+}_e = \sum_{f \neq e}\boS^+_{e,f}.
 \label{tot}
\end{equation}

\begin{figure*}
\includegraphics[width=\textwidth]{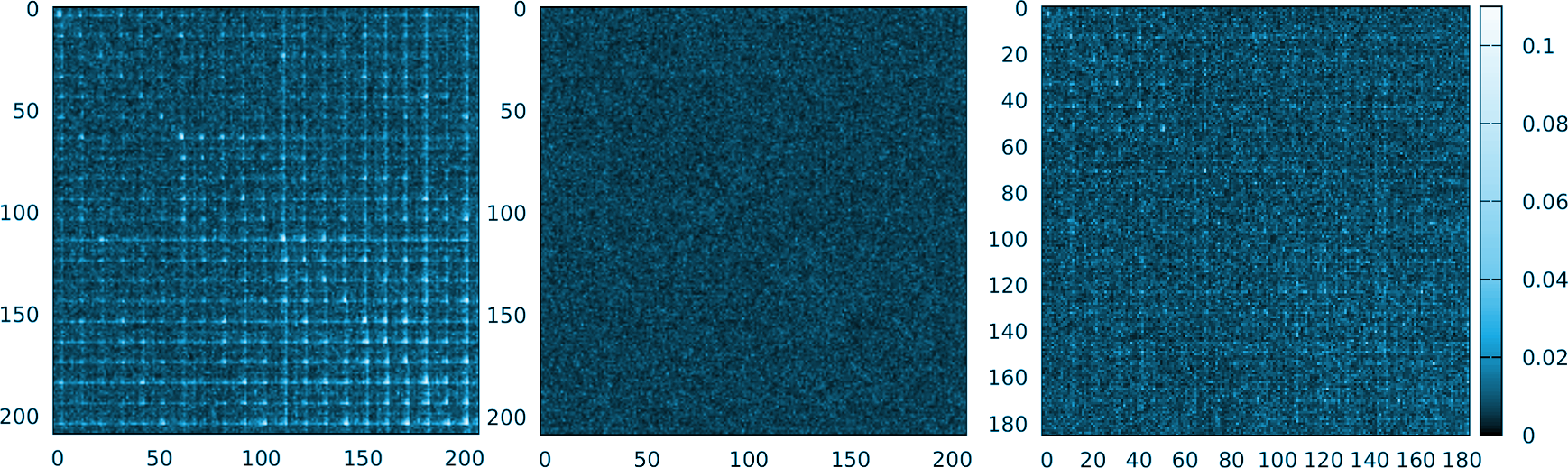} 
\caption{$\boS^+$for CFHTLenS, depicting which residual non-Gaussian correlations remain in the data set, after all Gaussian correlations were destroyed. Left: the original CFHTLenS data, displaying high levels of non-Gaussianity affecting $\xi_+$. The datavector is ordered as in the public CFHTLenS data products, most of the conspicuous data points are $\xi_+$ on angular scales of $\approx$ 35 arc minutes. Middle: The white-noise matrix of what CFHTLenS should look like if it were a Gaussian dataset. Right: The cleaned CFHTLenS data set, obtained by excluding the data points of highest non-Gaussianity. These are marked with asterisks in Table \ref{Tab}. The cleaning removes the strongest non-Gaussianities, but the cleaned dataset is still not entirely compatible with a multivariate Gaussian data vector, as indicated by the remaining structures.}
\label{CFHT}
\end{figure*}

We repeat the above procedure for the multiplication and division. From histogramming
\begin{equation}
 p_i^{e,f} = x_i^e*x_i^f,
\end{equation}
we build up the multiplicative trans-covariance matrix
\begin{equation}
\frac{1}{B} \sum_{b = 1}^B [ \calH_b(p_i^{e,f}) - \calP  ]^2 =: \boS^*_{e,f}.
 \label{Sstar}
\end{equation}
From histogramming the ratios
\begin{equation}
 r_i^{e,f} = x_i^e/x_i^f,
\end{equation}
we combine the individual MISE errors into the trans-covariance matrix
\begin{equation}
\frac{1}{B} \sum_{b = 1}^B [ \calH_b(r_i^{e,f}) - \calC  ]^2 =: \boS^\div_{e,f}.
 \label{Sdiv}
\end{equation}
From the matrices $\boS^*_{e,f}$   and $\boS^\div_{e,f}$ then follow $\epsilon^{\rm tot,\div}$,  $\epsilon^{\rm tot,*}$.

In short, our tests first destroy the Gaussian correlations between data points, and then measure which residual non-Gaussian correlations remain. The test asesses whether or not the whitened datapoints follow the correct distributions for the sum, product and ratio of Gaussian random variables and the mismatch between the expected and the observed distribution is used as a summary statistic for the level of non-Gaussianity. The larger the mismatch, the stronger the non-Gaussian correlations between the two datapoints. In the next subsection we study the sensitivity of the tests, before applying them in Sec.~\ref{Sec:Sec2} to CFHTLenS \citep{Heymans}.

\subsection{Sensitivity of the tests}
The proposed tests derive their sensitvity from measuring the MISE error of a histogrammed distribution $\calH(x)$ with respect to a known distribution $f(x)$. It is clear that for infinitely many random simulations, the histograms will be noise-free and the sensitivity of the test will then increase if the number of histogram bins is increased. However, given a finite number of simulations, the histogram bins will be subject to shot noise which limits the sensitivity of the tests. These imperfections must be accounted for. 

Before analyzing a potentially non-Gaussian dataset, we therefore implemented calibration runs. These calibration runs compute the covariance matrix of the potentially non-Gaussian dataset, and then draw $N_s$ truly Gaussianly distributed calibration datasets with the same covariance matrix. These calibration datasets are then used to calculate calibration matrices $\sfS^{+,*,\div}$, for $N_s$ Gaussian data sets. This allows to study the effects of shotnoise. 

We found that sometimes the shotnoise in the peak bins of the histograms can introduce tartan-like features in the trans-covariance matrices, even for Gaussianly distributed data. This is more often the case for the sharply peaked Cauchy distribution and the also sharply-peaked product-distribution $\calP$. For the comparably wide Gaussian distribution, this is barely an issue. Nonetheless, we always implemented 10 or 20 such calibration runs per setup of the pipeline, and then optimized the chosen number of histogram bins such that the shotnoise-induced structures are minimized. In most cases, they disappeared completely for an optimal number of histogram bins. Non-Gaussianity then produces additional and typically also visually very different structures, that clearly stand out beyond the shotnoise.

After these calibration runs, we left the pipeline untouched and tested the actual datasets. In the following we will always display the shotnoise trans-covariance matrices of Gaussian calibration runs side by side with the trans-covariance matrices for the tested non-Gaussian datasets.

To further study the sensitivity of the tests, we generated 220 Fourier modes $f(\fatk_i)$ of a non-Gaussian density field, produced by adding a squared term to the Gaussian field
\begin{equation}
 \Phi_{NG} = \Phi_G + f_{\rm nl}(\Phi_G^2 - \langle \Phi_G^2 \rangle ).
 \label{fnl}
\end{equation}
This is the so-called local model of non-Gaussianity, as familiar from the Planck analyses \citep{PlanckNG}. The scalar $f_{\rm nl}$ measures the amplitude of the non-Gaussianity. In Fig.~\ref{fnlplot} we display the trans-covariance matrices for 1000 simulations, with $f_{\rm nl} = 0.5$ and $\Phi_G$ drawn from a Gaussian of unit variance\footnote{Note that local non-Gaussianity in the Cosmic Microwave Background (CMB) applies the $f_{nl}$ transformation to the potential field, which has a variance of $\sim 10^{-10}$, so a $f_{nl}=1$ value here corresponds to a CMB value of $\sim 10^5$}. 
The amplitude $f_{\rm nl}$ was purposefully chosen to be large, in order to display the effect of the non-Gaussianity on the trans-covariance matrices. The more simulations are available, the sooner do the trans-covariance matrices detect non-Gaussian signatures beyond the shotnoise. This demonstrates that our setup is general and can in principle also detect very faint sources of non-Gaussianity, if only enough simulations are available. Here, we chose the MISE as a distance measure between the sampled histogram and the expected distribution function. 
The chosen distance measure is to a certain extent arbitrary, as long as the Gaussian calibration runs and the analyzed data set are treated in precisely the same way, such that shot noise in the histogram bins and non-Gaussianity can be told apart, independent of the selected distance measure. The trans-covariance matrices for addition and multiplication in Fig.~\ref{fnlplot} are symmetric matrices because the ordering does not matter. The matrix $\sfS^{\div}$ is however not symmetric, since $a/b \neq b/a$.

\newpage

\begin{figure*}
\includegraphics[width=0.9\textwidth]{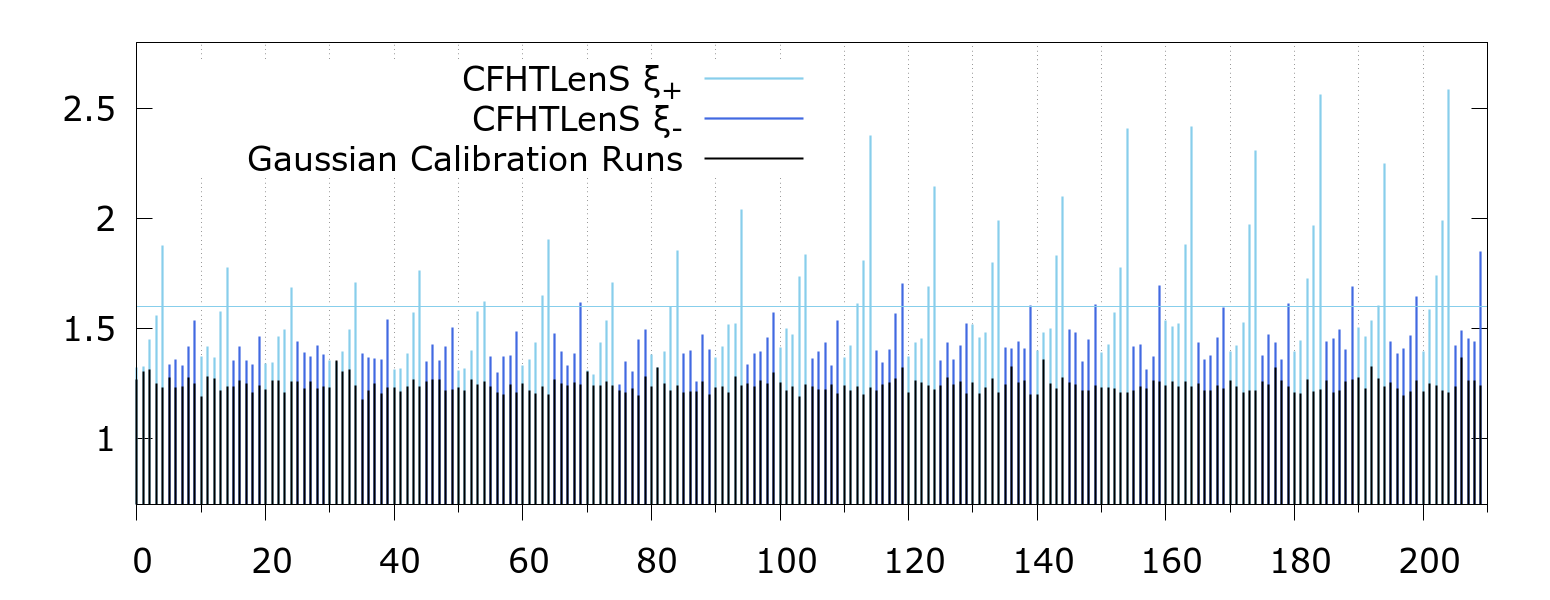} 
\caption{Total non-Gaussianity $\epsilon^{\rm tot,+}$ from Eq.~(\ref{tot}) in the 210 data points of CFHTLenS. Depicted in black are spurious traces of fake non-Gaussianity in a limited number of Gaussian random samples. The non-Gaussianities in blue refer to the CFHTLenS data, which clearly exhibit between 20-100\% non-Gaussian contaminations.}
\label{Lawn}
\end{figure*}

\section{Non-Gaussianities in CFHTLenS}
\label{Sec:Sec2}
As an example application, we consider the CFHTLenS cosmic shear survey. CFHTLenS is to date the deepest weak-lensing survey whose data- and simulation-products are publicly available. The main analyses of this survey use the angular correlation functions
\begin{equation}
  \xi_\pm(\theta) = \int \int J_{0,4}(l\theta) \frac{q_i(\chi) q_j(\chi)}{a^2(\chi)} \hat{P}_m\left(\frac{l}{\chi},\chi\right)  \mathd l\ \mathd \chi,
\end{equation}
where $J_{0,4}$ are Bessel functions of the first kind. The estimator $\xi_+$ filters with $J_0$, and $\xi_-$ filters with $J_4$. The $q_i(\chi)$ are weight functions which parameterize the lensing efficiency as a function of comoving distance $\chi$. The cosmological matter powerspectrum $P_m\left(\frac{l}{\chi}\right)$ is evaluated at wavemode $l/\chi$ at the cosmic epoch given by $\chi$. We refer the reader to \citet{Heymans, BartelSchneid} for a detailed introduction to weak lensing and its implementation in CFHTLenS.

\begin{table}
	\centering
	\caption{A list of data points from CFHTLenS, which are most contaminated with non-Gaussian correlations. The points are identified by their angle of separation $\theta$, their redshift bin combination where the lower bin is listed first in the tuple $(z_{\rm  low},z_{\rm high})$ and type of correlation function $\xi_+$ or $\xi_-$. The first column lists the datapoints position in the original CFHTLenS data vector, counted from zero, in the last column $\epsilon^+$ gives the total non-Gaussian contamination of the datapoint, according to Eq.~(\ref{tot}) and Fig.~(\ref{Lawn}). The points marked with asterisks cause the prominent stripes in Fig.~\ref{CFHT}.}
	\label{Tab}
	\begin{tabular}{lcccr lcccr} 
		\hline
		\# & $\theta$ & $z$-bins & $\xi_\pm$ & $\epsilon^{\rm tot,+}$ \\
		\hline 
		$204^*$ & 35.06 & (6,6)  &$\xi_+$& 2.59\\
		$184^*$ & 35.07 & (5,5)  &$\xi_+$& 2.57\\
		$164^*$ & 35.08 & (4,5)  &$\xi_+$& 2.42\\
		$154^*$ & 35.10 & (4,4)  &$\xi_+$& 2.41\\
		$114^*$ & 35.07 & (3,3)  &$\xi_+$& 2.38 \\
		$174^*$ & 35.08 & (4,6)  &$\xi_+$& 2.31\\
		$194^*$ & 35.07 & (5,6)  &$\xi_+$& 2.25\\
		$124^*$ & 35.09 & (3,4)  &$\xi_+$& 2.15 \\
		$144^*$ & 35.07 & (3,6)  &$\xi_+$& 2.10 \\
		$94^*$ & 34.99 & (2,5)   &$\xi_+$& 2.04 \\
		$134^*$ & 35.07 & (3,5)  &$\xi_+$& 1.99 \\
		$203^*$ & 16.61 & (6,6)  &$\xi_+$& 1.99\\
		$173^*$ & 16.63 & (4,6)  &$\xi_+$& 1.97\\
		$183^*$ & 16.60 & (5,5)  &$\xi_+$& 1.97\\
		$64^*$ & 34.89 & (2,2)   &$\xi_+$& 1.90 \\
		163 & 16.62 & (4,5)  &$\xi_+$& 1.88\\
		$4^*$ & 35.03  & (1,1)   &$\xi_+$& 1.87\\
		$84^*$ & 35.01 & (2,4)   &$\xi_+$& 1.86 \\
		209 & 35.06 & (6,6)  &$\xi_-$& 1.85\\
		$104^*$ & 34.98 & (2,6)  &$\xi_+$& 1.84 \\
		143 & 16.63 & (3,6)  &$\xi_+$& 1.83 \\
		$113^*$ & 16.62 & (3,3)  &$\xi_+$& 1.81 \\
		133 & 16.64 & (3,5)  &$\xi_+$& 1.80 \\
		14 & 34.94 & (1,2)   &$\xi_+$& 1.78\\
		44 & 35.03 & (1,5)   &$\xi_+$& 1.77 \\ 
		$153^*$ & 16.64 & (4,4)  &$\xi_+$& 1.76\\
		103 & 16.65 & (2,6)  &$\xi_+$& 1.74 \\
		$202^*$ & 7.70  & (6,6)  &$\xi_+$& 1.74\\
		182 & 7.70  & (5,5)  &$\xi_+$& 1.73\\
		34 & 35.06 & (1,4)   &$\xi_+$& 1.71\\
		119 & 35.07 & (3,3)  &$\xi_-$& 1.71 \\
		74 & 34.98 & (2,3)   &$\xi_+$& 1.71\\ 
		123 & 16.63 & (3,4)  &$\xi_+$& 1.70 \\
		159 & 35.10 & (4,4)  &$\xi_-$& 1.70\\
		24 & 35.04 & (1,3)   &$\xi_+$& 1.69\\   
		189 & 35.07 & (5,5)  &$\xi_-$& 1.69\\
		$63^*$ & 16.64 & (2,2)   &$\xi_+$& 1.65 \\
		199 & 35.07 & (5,6)  &$\xi_-$& 1.65\\
		54 & 35.04 & (1,6)   &$\xi_+$& 1.62 \\	
		69 & 34.89 & (2,2)   &$\xi_-$& 1.62\\
		179 & 35.08 & (4,6)  &$\xi_-$& 1.61\\
		$193^*$ & 16.61 & (5,6)  &$\xi_+$& 1.61 \\
		$83^*$ & 16.65 & (2,4)   &$\xi_+$& 1.60 \\ 
		\hline
	\end{tabular}
\end{table}

Of concern here is the standard approach in weak lensing analyses, where the estimated correlation functions $\xi_\pm(\theta)$ are assumed to follow a Gaussian likelihood, such that for parameter inference, only a covariance matrix has to be provided along with the datavector. Estimating covariance matrices for the modern sky surveys proves however to be a formidable task, posing utmost demands on numerical simulations \citep{Teyssier} and physically motivated modelling techniques \citep{KiDS}. Before these challenges are to be adressed by the cosmological community, it is certainly an interesting question to wonder whether calculating a covariance matrix is actually the task to be executed. On the on hand, if the likelihood turns out to be non-Gaussian, then computing a covariance matrix at arbitrary precision will always be insufficient. On the other hand, many non-Gaussian likelihoods do not even require the provision of a covariance matrix. Concerning weak-lensing, we will show in the following that a Gaussian likelihood is indeed only an approximation to the yet unknown true likelihood.

We applied our method to the Clone simulations \citep{Clone1, Clone2, Clone3} for the tomographic analysis of the CFHTLenS data set as presented in \citet{Heymans}. These provide 1656 semi-independent simulations for the 210 datapoints of CFHTLenS. Using these simulations, our trans-covariance matrices clearly detect non-Gaussian correlations between various data points. The $\boS^+$ matrix for CFHTLenS is depicted in the left panel of Fig.~\ref{CFHT}, but the non-Gaussian couplings are also identified in $\boS^*$ and $\boS^\div$. The displayed trans-covariance matrix has the same structure as the CFHTLenS covariance matrix. To be precise, the data vector on the axes of the matrix plot is ordered as in the publicly available data products, meaning all five angular measurements of $\xi_+$ are followed by all five angular measurements of $\xi_-$, which then repeats for increasing combinations of redshift bins. 

Somewhat surprisingly, most often affected by non-Gaussian correlations are the largest angular scales, rather than the smallest where non-linear structure growth would have been the likely origin for the non-Gaussianty. The most prominent stripes in $\boS^+$ are caused by $\xi_+$ on angular scales of about 35 arc minutes. The correlation function $\xi_-$ is less affected by non-Gaussianities. The most contaminated data points are given in Tab.~\ref{Tab}. 

\subsection{Strength of the non-Gaussianities}
The prominent stripes in the trans-covariance matrix Fig.~\ref{CFHT} already flag the most contaminated data points. The contamination per data point is made more quantitative by computing the total per row or column, as given in Eq.~(\ref{tot}). These total contamination levels are listed in Table \ref{Tab}, and depicted in Fig.~\ref{Lawn} for all data points. The black impulses in Fig.~\ref{Lawn} indicate the total of the MISE for Gaussian random variables of the same covariance matrix as CFHTLenS. The floor provided by these black lines is to be interpreted as a threshold below which non-Gaussianities cannot be detected anymore, since the limited number of simulations erase that information. Depicted in blue shades are then the contaminations in CFHTLenS, which clearly stand out above the shotnoise. 

The magnitude of these non-Gaussianities is best understood by studying the sensitivity of the tests for the local $f_{\rm nl}$ model: $f_{\rm nl}$ is a scalar that represents the amplitude of non-Gaussianity. For example, the bispectrum of the local model for non-Gaussianity is proportional to $f_{\rm nl}$. From Eq.~(\ref{fnl}) it is also clear that $f_{\rm nl}$ indicates the fractional contribution of the squared non-Gaussian fields $\Phi^2_G - \langle \Phi^2_G\rangle$ to the total field value. An $f_{\rm nl}$ value of unity then indicates that the non-Gaussian and the Gaussian field contribute equal power to the total field. We hence use the minimally detectable $f_{\rm nl}$ as a `proxy' for the lower bound on detected generic non-Gaussianities.

For the 1656 CFHTLenS simulations, and 210 data points, the trans-covariance matrices succeed in detecting local non-Gaussianities of $f_{\rm nl} \geq 0.3$. We therefore conclude that the non-Gaussianities detected in CFHTLenS must be of at least comparable amplitude, meaning they contribute a faction of at least $0.3$ to the total statistical uncertainties in $\xi_+$ and $\xi_-$. A Gaussian approximation to the distribution of $\xi_+$ and $\xi_-$ will hence ignore about 30\% of the statistical correlations between the datapoints.

The trans-covariance matrices so far employed measure non-Gaussianities in joint distributions of two data points. In the next subsection, we will complement these findings by an analysis of the individual distributions, which also exhibit non-Gaussianities. In the remainder of the paper, we will present a preliminary study of the impact that these non-Gaussianities have on parameter constraints.

\subsection{Individual distributions}
\begin{figure*}
\includegraphics[width=0.95\textwidth]{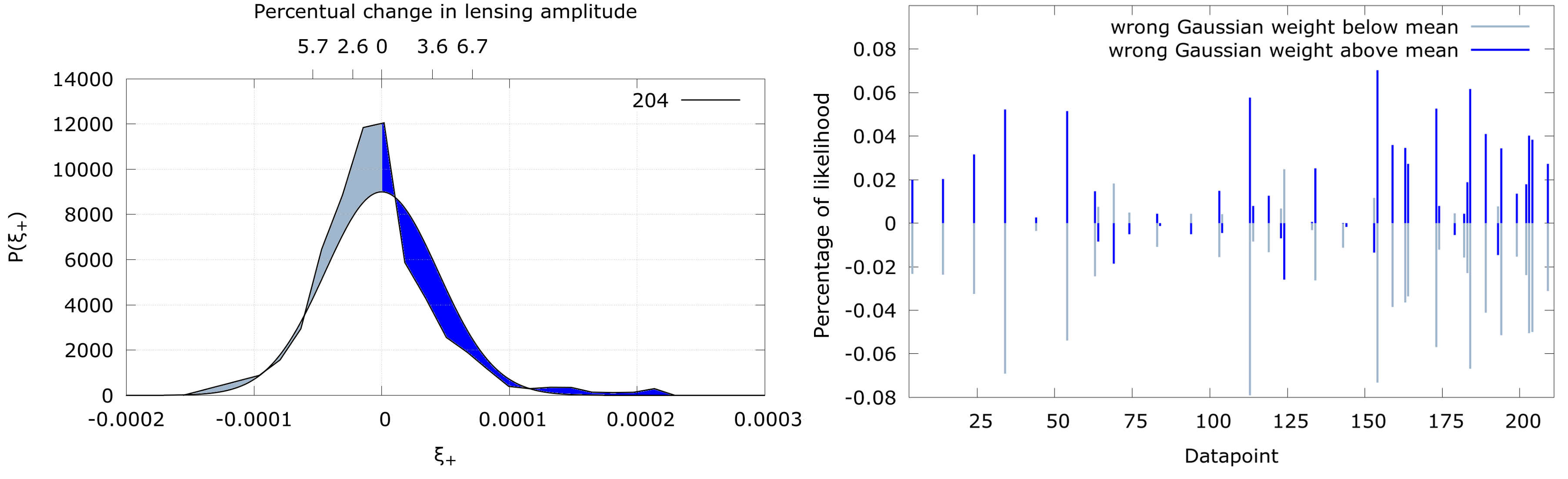} 
\caption{Comparison between the distribution function of individual data points, in order to complement the detected non-Gaussianities in the joint distributions. Left: comparison between the actual distribution and its Gaussian approximation as it currently enters weak-lensing analyses. The panel refers to datapoint 204, see Table \ref{Tab} for details. The same pattern of skewness is seen for all other contaminated datapoints as well. Right: Depiction of the wrong weight assignments for all datapoints from Table \ref{Tab}. The Gaussian clearly puts too little weight below the mean, and too much weight above the mean. The colour and magnitude of the vertical bars indicate the areas between the two probability density functions as plotted on the left.}
\label{Weights}
\end{figure*}

In the previous section, we had focused on joint distributions of all possible pairs of data points. Here we investigate the distribution of the individual data points.

In Fig.~\ref{Weights} we plot the mean-subtracted sampling distribution from the Clone simulations, for the 204th data point, which is $\xi_+(\theta = 35 {\rm arcmin})$ from the redshift bins $(6,6)$. The figure reveals that the sampling distribution displays skewness towards values lower than the mean. This left-skewness is then compensated by a tail stretching to high amplitudes of $\xi_+$. The skewed shape of the distribution is generic and also arises for the other contaminated data points. Since the peaks of these distributions are systematically at lower values than their means, it is then most likely that a weak-lensing data set contains many data points which systematically indicate a low lensing amplitude. This is a natural consequence of the left-skewness and does not indicate any problem of the data. It does however indicate that the correct analysis of weak-lensing data should employ a non-Gaussian likelihood. Otherwise, parameter constraints derived from the weak-lensing amplituded will be biased low. Since the weak-lensing amplitude increases with the matter density $\Omega_m$ and the variance $\sigma^2_8$ of the matter powerspectrum, this has the potential to explain the currently seen discrepancy between Planck, CFHTLenS and KiDS. 

To further demonstrate the point, Fig.~\ref{Weights} also includes the Gaussian approximation to the likelihood, as currently used in weak-lensing analysis. The upper x-axis of the left panel shows the percentage by which the weak-lensing amplitude has to be increased or decreased, in order to shift an estimated weak-lensing signal through the histogram. We see that the left-skewness leads to a most likely lensing amplitude that is about $5\%$ lower than the mean. The same holds true for the other contaminated datapoints. The Gaussian approximation of the likelihood, which is centered on the mean, instead of the peak, therefore systematically underestimates the probability of a dataset displaying lensing amplitudes lower than the mean. 

We quantify the systematically misassigned probabilities as follows. Let $\mathcal{G}(\xi_i)$ denote the Gaussian approximation to the true distribution $\mathcal{S}(\xi_i)$ of the $i$th data point of a weak lensing correlation function.

The probability that the Gaussian approximation assignes incorrectly to values lower than the mean is then 
\begin{equation}
W_L =  \int_{-\infty}^{\bar{\xi_i}} \left[ \mathcal{G}(\xi_i) - \mathcal{S}(\xi_i) \right] \mathd \xi_i.
\end{equation}
This is indicated as bright blue area in the left of Fig.~\ref{Weights}. The incorrectly assigned likelihood to values above the mean is likewise
\begin{equation}
W_A =  \int_{\bar{\xi_i}}^\infty \left[ \mathcal{G}(\xi_i) - \mathcal{S}(\xi_i)  \right] \mathd \xi_i,
\end{equation}
which corresponds to the dark-blue area in Fig.~\ref{Weights}. These incorrectly assigned probabilities $W_+$ and $W_-$ are plotted in the right of Fig.~\ref{Weights} for all contaminated datapoints of Table \ref{Tab}. This reveals that the Gaussian likelihood approximation to all contaminated datapoints systematically underestimates the probability of low lensing-amplitudes by about 2-10\% (bright blue). Likewise, it overestimates the likelihood of high lensing-amplitudes (dark blue). In order to reconcile the currently measured lensing amplitudes of CFHTLenS with the lensing amplitude predicted from Planck, an approximate $10\%$ increase in the amplitude is needed. Our study so far indicates that the preference of low lensing amplitudes due to non-Gaussianities has the right order of magnitude and the right sign in order to account for this discrepancy. A final statement can however only be made upon availability of the correct non-Gaussian likelihood.

\section{Effect on parameter constraints}
\label{Sec:Sec3}
\label{DiscAgree}

As demonstrated in the previous section, the correlations between various data points of CFHTLenS give rise to non-Gaussianities at a 30\% level according to our definition. Here, we present a preliminary study of how these non-Gaussianities might impact parameter constraints, by excluding the most contaminated data points from the likelihood. However, as essentially the entire CFHTLenS dataset is contaminated (see Fig.~\ref{Lawn}), such exclusions are clearly a suboptimal strategy. We nonetheless report our findings as intermediate results and postpone an update to a non-Gaussian likelihood to future work. 

\begin{figure*}
\includegraphics[width=0.9\textwidth]{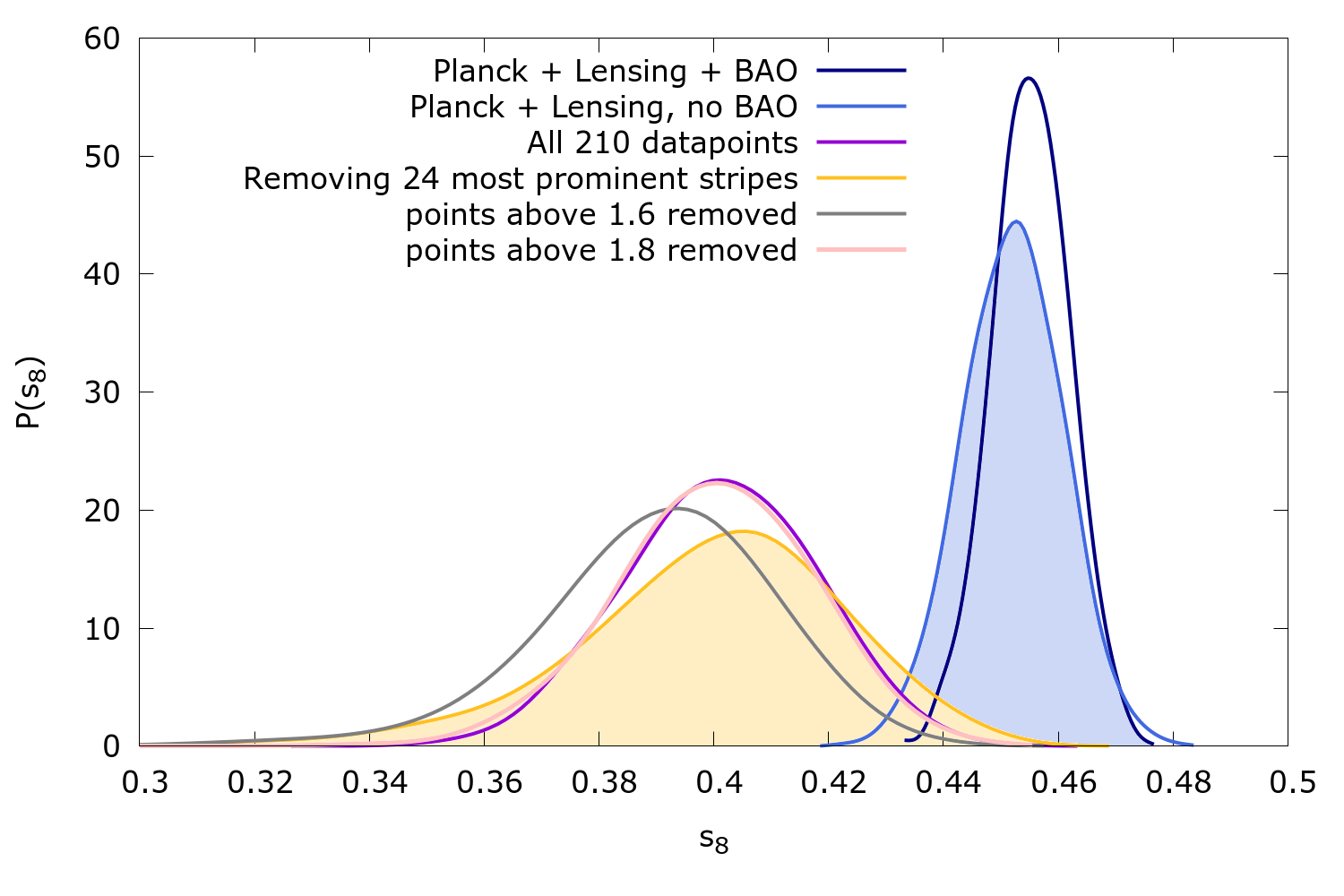} 
\caption{Comparison of the $s_8$ posteriors, as gained from different cuts to the CFHTLenS data, and two Planck analyses. The open (filled) blue contours indicate Planck 2015 data with (and without) BAO, both times including lensing. The two pink shades correspond to CFHTLenS, once the full dataset of 210 points, and once with 23 points of non-Gaussian contaminations above 1.8 removed, see Table \ref{Tab}. These agree remarkably well, even though 10\% of the dataset were excluded. Yellow indicates the removal of 24 points which cause the prominent stripes as depicted by Fig.~\ref{CFHT}. These points are marked by asterisks in Table \ref{Tab}. Grey depicts the removal of all 43 points in Table \ref{Tab} with a contamination level above 1.6. }
\label{s8}
\end{figure*}

We use the publicly available Cosmosis \citep{Cosmosis} pipeline to reanalyze the CFHTLenS data, subjected to multiple cuts. We have analysed the 6-bin tomographic data from the CFHTLenS weak lensing survey, as described in \citet{Heymans}. A standard flat LCDM model is assumed, with cosmological parameters $\Omega_m$, $\sigma_8$, $n_s$, $\Omega_b$, and $h$, and additionally a linear intrinsic alignment model is included, with amplitude $a$ as an additional parameter. Three photometric redshift bias parameters, three shear calibration parameters and the optical depth to reionization are kept fixed. Dark energy is modelled as a usual cosmological constant. The power spectrum is computed with CAMB and Halofit, and then log-linear extrapolated to high $k$ \citep{HaloCamb1, HaloCamb2, HaloCamb3}. The basic redshift distribution is read in in tabular form. We ran four analyses, once using all $210$ data points of the original dataset, once excluding the 24 datapoints marked by asterisks in Table \ref{Tab}, which cause the stripy features in Fig.~\ref{CFHT}, once excluding all datapoints with a total error of $\epsilon^{\rm tot,+} \geq 1.6$, and once for $\epsilon^{\rm tot,+} \geq 1.8$. Table \ref{Tab} lists these datapoints. Fig.~\ref{s8} depicts the posterior on $s_8 = \sqrt{\Omega_m}\sigma_8$ for these runs. Also included are the constraints on $s_8$ from two Planck analyses: The Planck datasets here used are both from the 2015 analysis of the temperature-temperature (TT) spectrum, the temperature and E-mode spectrum (TE), and the E-mode spectrum (EE), together with lensing and low TEB. They differ in whether or not also data from Baryonic Acoustic Oscillations (BAO) are included. These are two of the most constraining probe combinations for $\Omega_m$ and $\sigma_8$.

The posteriors in Fig.~\ref{s8} reveal that excluding the 24 points (11\% of the dataset) from Table~\ref{Tab}, which cause the stripes in Fig.~\ref{CFHT} shifts the posterior of $s_8$ towards those from Planck. However, excluding the 23 points (10\% of the dataset) of contamination levels above 1.8 leads to virtually the same constraints on $s_8$ as the full dataset. Excluding all 43 datapoints (20\% of the dataset) with non-Gaussian contaminations above a level of $1.6$ shifts the posterior away from Planck. The posteriors broaden only insignificantly by the exclusion of the datapoints, illustrating that about 20\% of the dataset can allegedly be excluded without loss of precision, while the overlap with Planck can be in- and decreased depending on which points are in- and excluded. The simple in- and exclusion of datapoints does not yet, account for non-Gaussian correlations in the remaining data points, such that final conclusions have to be deferred until the correct non-Gaussian likelihood is available.  We see that there is no clear signature that removal of the most non-Gaussian data points shifts the posterior significantly.  However, there is no natural cut on the non-Gaussianity that we can usefully apply, since all of the data points show significant deviations from gaussian expectations.  We leave deeper investigation to further work.

\section{Discussion}
This paper addressed the question of whether cosmological observables follow Gaussian likelihoods, and how generic deviations from Gaussianity can be detected. One way of detecting non-Gaussianities in a likelihood is to generate many independent synthetic samples of the data vector. In this paper, we have focussed on CFHTLenS, for which 1656 simulations for its 210 dimensional data vector are available. 1656 random samples of a 210 dimensional data vector are however not sufficient to map out the data vector's likelihood in a conventional way. To detect non-Gaussianities in the data vector's distribution, we therefore implemented a novel unconventional testing pipeline, which essentially proceeds as follows: first the Gaussian covariances between the elements of a data vector are destroyed. Then, the tests measure whether any residual non-Gaussian correlations remain. The tests then flag all pairs of data points which display significant non-Gaussian correlations. If such points exist, then even an arbitrarily precise covariance matrix will be insufficient for describing the statistical dependencies between the two data points. 

We detected an approximate 30\% contribution of non-Gaussianities to the statistical uncertainties of the CFHTLenS data set, affecting especially $\xi_+$ on the largest angular scales. The skewed non-Gaussian probability density functions of the individual datapoints furthermore suggest that weak lensing data sets are likely to display low lensing amplitudes, without this indicating any hidden systematics in the data. This systematic preference of low lensing amplitudes caused by the skewness has the same order of magnitude and the right sign to explain the tension between the CFHTLenS, the KiDS and the Planck data. In a preliminary analysis that simply excludes the most non-Gaussian data points, we found shifts in the posterior of $s_8$, however no clear pattern emerged: we were able to exclude about 20\% of the data without losing statistical precision, yet the exclusion of datapoints resulted in shifts of the posterior. This indicates that currently, statistical uncertainties on parameters are still dominating the entire uncertainty. 

A simple exclusion of data points obviously does not fully address the issue of non-Gaussian correlations. A further more detailed study into the origin and effects of these detected non-Gaussianities is under way. Currently, two scenarios for the origin of the non-Gaussianities seem likely: they could arise because the light rays of distant galaxies propagate through non-Gaussian matter fields before they are detected. They could also arise in Gaussian matter fields, since the weak-lensing correlation functions are quadratic forms and are therefore expected to exhibit $\chi^2$-like deformations of the likelihood. The computation and validation of such a non-Gaussian likelihood involves a comparison with tailor-made simulations, and is therefore also postponed to a future paper.

As the detected non-Gaussianities primarily affect large angular scales, future weak lensing surveys like Euclid and LSST can be expected to be affected by this issue. We recommend running the described tests when estimating the covariance matrix for any survey from simulations. If the tests indicate the presence of non-Gaussianity, then estimating a covariance matrix alone may be insufficient, and parameter constraints derived from assuming a Gaussian likelihood will be biased.

Addendum: During peer review, the Dark Energy Survey (DES) published the cosmology results of their 1-year data \citep{Troxel} and we were repeatedly asked to comment. We can to-date not yet estimate the extent to which DES is affected by the non-Gaussianities described here, as DES uses an analytically-approximated covariance matrix. A comparison with numerically-estimated covariances has however been announced by the DES consortium (MacCrann et al. in prep). Further work on our side has however demonstrated that non-Gaussianities in shear correlation functions are generic to weak lensing analyses, and KiDS and LSST are indeed affected (Sellentin, Heymans et al. in prep.)

\section{Acknowledgements}
We thank Catherine Heymans, Joe Zuntz, Linda Blot, Tom Kitching, Matthias Bartelmann and Shahab Joudaki for support, provision of simulation products and answering detailed questions. We thank an anonymous referee for peer-review.
E.S. is funded by a DAAD research fellowship of the German Academic Exchange Service. Cosmosis can be found at https://bitbucket.org/joezuntz/cosmosis/wiki/Home  .
Computations for the [Clone] N-body simulations were performed on the TCS supercomputer at the SciNet HPC Consortium. 
SciNet is funded by: the Canada Foundation for Innovation under the auspices of Compute Canada; 
the Government of Ontario; Ontario Research Fund - Research Excellence; and the University of Toronto.

\bibliographystyle{mn2e}
\bibliography{TDist}

\label{lastpage} 
\bsp 
\end{document}